\documentstyle[preprint,aps,eqsecnum]{revtex}
\renewcommand{\i}{\textrm{i}}
\renewcommand{\d}{\textrm{d}}

\tighten
\begin{document}
\title{Tensionless string in the notoph background}
\author{{\it K~Ilienko}\thanks{E-mail address: ilienko@maths.ox.ac.uk \& 
        kost@ire.kharkov.ua},
         \\ Balliol College and Mathematical Institute, 
         \\ 24--29 St Giles', Oxford OX1 3LB, UK 
	 \\ and \\ 
	 Usikov Institute for Radiophysics and Electronics, \\
	 12~Acad~Proscura Street, Kharkiv 310085, Ukraine \\[4ex]
	{\it A~A~Zheltukhin}\thanks{E-mail address: 
	 zheltukhin@kipt.kharkov.ua}, \\ 
         Kharkiv Institute for Physics and Technology, \\ 1~Akademichna 
	 Street, Kharkiv 310108, Ukraine 
         \\ and \\ 
         Institute of Theoretical Physics, University of Stockholm, \\
         Box~6730, S--11385, Stockholm, Sweeden} 
%\date{August 14, 1998}
\maketitle
%------------------------------------------------------------------------------
\begin{abstract}
We study the interaction between a tensionless (null) string and an
antisymmetric background field $B_{ab}$ using a 2-component spinor
formalism. A geometric condition for the absence of such an interaction is
formulated. We show that only one gauge-invariant degree of freedom
of the field $B_{ab}$ does not satisfy this condition.
Identification of this degree of freedom with the notoph
field $\phi$ of Ogievetskii-Polubarinov-Kalb-Ramond
is suggested. Application of a 2-component spinor formalism allows one a
reduction of the complete system of non-linear partial differential
equations and constraints governing the interacting null string
dynamics to a system of linear differential equations for the basis
spinors of the spin-frame.  We find that total effect of the interaction
is contained in a single derivation coefficient which is
identified with the notoph field.
\end{abstract}
%------------------------------------------------------------------------------
%\vspace{-1.5cm}
%\pacs{11.25}
%\vspace{-1cm}
%------------------------------------------------------------------------------
\newpage
\section{Introduction}
Recently, there has been considerable interest in the investigation of the
string and null string equations of motion in different backgrounds
of external fields and in curved spacetimes
(e.g. \cite{Veneziano,deVega-Sanchez,Antoniadis} and references
therein). Finding the exact solutions for the equations of motion in
such
systems is a rather difficult task, mainly due to the non-linear
character of the equations of motion, so it seems interesting to study
those situations in which these equations possess exact solutions. It
is known that the equations of motion for strings in 4D~Minkowski
spacetime, null strings in some backgrounds and for particular types
of curved spacetimes can be solved exactly
\cite{deVega-Sanchez,Antoniadis,Roshchupkin-Zheltukhin,Kar(1-2),Dabrowski-Larsen}.
Being a zero-tension limit of strings \cite{Schild}, null strings
possess simpler equations of motion than those of strings and can be
considered as a zero approximation with the string tension as the
perturbation
parameter \cite{Zheltukhin1,deVega-Giannakis-Nicolaidis,Lousto-Sanchez}.
This explains our current interest in this problem.

Geometrically worldsheets of null strings are lightlike (null)
2-surfaces which generalize worldlines of massless particles
\cite{Schild,Karlhede-Lindstrom,Lizzi,Zheltukhin(2-3)}. By
convention, null string interactions with various fields can be analysed
into two types. To the first type we attribute the interactions which
violate the lightlike character of the worldsheets and can lead
to generation of non-zero tension
\cite{Zheltukhin4,Town1,Has-Lin,De-Ga-Ho-So,Bandos-Zheltukhin0} . All
other interactions fall into the second type. In particular, null string
interactions with antisymmetric background tensor fields are of the
second type.
 As shown in \cite{Zheltukhin4}, 2-component spinors of 3D~Minkowski
spacetime provide a particularly convenient framework for studying the
general solutions of null string equations of motion in arbitrary
antisymmetric
background fields $B_{ab}(x)$. On the other hand, a convenience of
adopting 2-component spinor description for spacelike and timelike
strings in 4D~Minkowski spacetime was demonstrated in \cite{Hughston-Shaw}
and
for free null (super) strings and p-branes in
\cite{Bandos-Zheltukhin}. This formalism is based on the fundamental
results of Penrose's twistor programme \cite{SST} and is a part of the
so-called twistor approach in the theory of supersymmetric objects
\cite{(1-2)Vol-Zhe,3Vol-Zhe,Sor-Tk-Vol,Sor-Tk-Vol-Zhe,Ho-Town,Gal-Soc,Iva-Kap,Chi-Pa,Ber-Sez,Berkov}.
From this viewpoint the goal of this article is to study further
the utility of such an approach in describing exact solutions of non-linear
equations of motion for extended objects in external fields. The first signs
as to the efficiency of this approach were found in
\cite{Zheltukhin4} where the dynamics of null strings in
3D~Minkowski spacetime and null membranes in 4D~Minkowski spacetime in
background antisymmetric fields $B_{ab}(x)$ and $B_{abc}(x)$ was
investigated. It has been shown that the contribution of these fields into
the null string and null membrane equations of motion can be annihilated
by reparametrization of the corresponding null worldsheets.

On the other hand, from the point of view of gauge field theory for
$B_{ab}(x)$ in 3D~Minkowski spacetime and $B_{abc}(x)$ in 4D~Minkowski
spacetime these results are natural ones since $B_{ab}(x)$ and
$B_{abc}(x)$ are the pure gauge fields in the given dimensions
of spacetimes. Duality relations $B_{ab}=\varepsilon_{abc}\phi^{c}$ and
$B_{abc}=\varepsilon_{abcd}\phi^{d}$  allow exclusion of the vector
field $\phi^{a}$ from equations of motion by fixing gauge parameters
$\Lambda_a$ and $\Lambda_{ab}$  of the corresponding U(1) internal
group of gauge symmetry. Nevertheless, the non-trivial nature of
results \cite{Zheltukhin4} lie in the fact that they were
derived without using the gauge symmetry described above. Utilization
of worldsheet Virasoro symmetry together with new gauge
symmetries of the 2-component spinor formalism were sufficient for
elimination of these pure gauge fields. These symmetries allow
reduction of the null string and null membrane non-linear equations of 
motion to linear and exactly solvable ones. Taking a pragmatical point of 
view it could be advantageous to exclude all non-physical degrees of 
freedom of the external fields interacting with extended objects by
fixing gauge conditions corresponding to the internal symmetry group
from the very beginning. However, this leads to the Lorentz non-covariant
description which is disadvantageous on the quantum level.

Our investigation suggests that 2-component spinor formalism allows one to
simplify the non-linear dynamics of strings and membranes propagating in
external fields or curved spacetimes without breaking the Lorentz
covariance and internal gauge symmetry invariance of the description
explicitly. We hope that simplifications achieved in the framework of
such a description could provide the most relevant choice of effective
variables which would permit  integration of non-linear motion
equations for p-branes in various spacetime dimensions.

Assuming the correctness of the above-stated hypothesis we shall study
non-linear equations of motion of a tensionless string embedded into
the background of an antisymmetric field $B_{ab}(x)$ in 4D~Minkowski
spacetime. In this case the field $B_{ab}(x)$ has physical degrees of
freedom which cannot be eliminated by fixing of U(1) gauge
symmetry. In what follows we have to keep in mind that the results
discussed in this paper depend on the dimension of spacetime
\cite{Jensen-Lindstrom}.

We show below that the use of Virasoro
reparametrization symmetry and local symmetries inherent in the
2-component spinor formalism provide a gauge invariant possibility to
exclude all non-physical degrees of freedom of the gauge field $B_{ab}(x)$. 
It turns out that only one gauge-invariant physical degree of freedom 
influences the null string dynamics, and we identify this degree of
freedom with the notoph field of Ogievetskii-Polubarinov-Kalb-Ramond
\cite{Ogievetskii-Polubarinov,Kalb-Ramond}.

We also formulate a geometric condition for the absence of interaction
between
the tensionless string and background antisymmetric fields which is
equivalent to the condition that $B_{ab}(x)$ is a pure gauge field.
This condition consists in the vanishing of the scalar product of
the 4-vector velocity vector $\dot{x}^a$ with field strength $Q_a(x)$ of the
gauge
field $B_{ab}(x)$.

Similar to the case of the Lund-Regge geometrical approach
\cite{Lund-Regge,Omnes,Zheltukhin6,Zheltukhin(5-7),Zheltukhin8,Barbashov-Nesterenko1}
we show that null string equations of motion can be reduced to algebraic
conditions on the coefficients of decomposition (the so-called
derivation coefficients) of the first derivatives of the basis spinors of
a 2-component spin-frame (dyad) with respect to worldsheet coordinates
$\tau$ and $\sigma$. Then integrability conditions for the
representation of the first derivatives of $x^a (\tau,\sigma)$ with
respect to $\tau$ and $\sigma$ start playing the role of the dynamical
equations.

We establish that the effect of interaction with the antisymmetric field
is contained in one of the derivation coefficients of $\dot{o}^A$, and
we interpret it as the notoph field $\phi$. From the point of view
of the gauge field theory approach
\cite{Zheltukhin6,Zheltukhin(5-7),Zheltukhin8} these derivation
coefficients have physical meaning of 2D~Yang-Mills or Higgs
fields. Therefore the notoph field finds interpretation as an
additional component of 2D~Yang-Mills-Higgs field.

The resulting equations of motion of the null string in a background of
antisymmetric fields can be written as a system of linear second-order
partial differential equations for the basis
spinors of the dyad. This shows that the 2-component spinor
formalism can be successfully employed for the purpose of
linearization of non-linear equations of motion of tensionless strings
interacting with background antisymmetric gauge fields in 4D~Minkowski
spacetime.
%------------------------------------------------------------------------------
%\newpage
\section{Equations of motion}
Let us consider the action of a null string in an external
antisymmetric field $B_{ab}(x)$. In the spinor form it can be rewritten
as follows:
\begin{equation}
S = \int \Big[ \rho^\mu \partial_\mu x^{AA'}
o_A\bar{o}_{A'}-
\kappa\varepsilon^{\mu_1 \mu_2}
\partial_{\mu_1} x^{A_1 {A'}_1}
\partial_{\mu_2} x^{A_2 {A'}_2}
B_{A_1 A_2 {A'}_1 {A'}_2}(x)\Big]\d^2 \xi ,
\label{2.1}
\end{equation}
where $B_{A_1 A_2 {A'}_1 {A'}_2}$, $x^{A A'}$ and $\d^2 \xi$
represent, respectively, the field $B_{ab}$, the coordinates of the null string
$x^a (\tau ,\sigma ) $ and the area element $\d\tau \d\sigma$ of the
worldsheet. We assume that $\xi^{\mu} = (\tau ,\sigma)$ is a
smooth parametrization of the null-string worldsheet, choose
$\varepsilon^{\tau\sigma} =-\varepsilon^{\sigma\tau}=-1$,
$\eta_{ab}=\mathrm{diag}(+--\,-)$ and use the notation
$\partial_\mu = \partial / \partial \xi^\mu$.
The coordinates $\xi$ on the worldsheet are dimensionless and we
assume the worldsheet vector density $\rho^\mu (\xi)$ to have the
dimensions of inverse length. The density $\rho^\mu$ ensures
invariance of action (\ref{2.1}) under arbitrary non-degenerate
reparametrisations of the null string worldsheet
\cite{Bandos-Zheltukhin0}. Interaction constant $\kappa$ has the dimensions
of inverse length.  Spinor fields  $o^A$ and $\iota^A$ form a basis for
2-dimensional complex vector space and obey the normalization conditions
\begin{equation}
o_A\iota^A =\overline{(\bar{o}_{A'}\bar{\iota}^{A'})} = \chi, \,\,\,\,
o_Ao^A = \iota_A\iota^A = 0,
\label{2.2}
\end{equation}
where $\chi(x^{AA'})$ represents a possibility of rescaling of the
dyad element ${\iota}^A$ at each spacetime point
$x^{AA'}$. The relation
\begin{equation}
B_{A_1 {A'}_1 A_2 {A'}_2} =
B_{A_1 A_2 {A'}_1 {A'}_2} =
-B_{A_2 A_1 {A'}_2 {A'}_1}
\label{2.3}
\end{equation}
holds for the antisymmetric tensor field $B_{ab}$ and we introduce
$\partial_{AA'} = \partial / \partial x^{AA'}$,
$\dot x^{AA'}=\:\partial_\tau x^{AA'}$,
$\acute{x}^{AA'}=\:\partial_\sigma x^{AA'}$ and
\begin{equation}
3\partial_{[AA'}B_{A_1 A_2 {A'}_1 {A'}_2]}=
\partial_{AA'}B_{A_1 A_2 {A'}_1 {A'}_2}+
\partial_{A_1 {A'}_1}B_{A_2 A {A'}_2 A'}+
\partial_{A_2 {A'}_2}B_{A A_1 A' {A'}_1}.
\label{2.4}
\end{equation}

Variation of action (\ref{2.1}) results in the equations describing the
null string dynamics
\begin{equation}
\begin{array}{l}
\rho^\mu \partial_\mu x^{AA'} o_A =0, \\
\partial_{\mu} x^{AA'}o_A\bar{o}_{A'} = 0, \\
\partial_\mu (\rho^\mu o_A\bar{o}_{A'})+
3\kappa\varepsilon^{\mu_1 \mu_2}
\partial_{\mu_1} x^{A_1 {A'}_1}
\partial_{\mu_2} x^{A_2 {A'}_2}
\partial_{[AA'}B_{A_1 A_2 {A'}_1 {A'}_2]}=0,
\end{array}
\label{2.5}
\end{equation}
and complex conjugate of them.
The first equation in (\ref{2.5}) and its complex conjugate imply
\begin{equation}
\rho^\mu\partial_\mu x^{AA'} = e o^A \bar{o}^{A'},
\label{2.6}
\end{equation}
where $e(\xi)$ is an arbitrary real-valued function with
transformation properties of a scalar worldsheet density. Assuming that
$\rho^\tau$ is a nowhere-zero function we rewrite this equation as
\begin{equation}
\dot{x}^{AA'}=
\frac{e}{\rho^\tau} o^A \bar{o}^{A'}-
\frac{\rho^\sigma}{\rho^\tau}\acute{x}^{AA'}.
\label{2.7}
\end{equation}
Taking into account the second equation in (\ref{2.5}) we obtain
\begin{equation}
\acute{x}^{AA'}o_A\bar{o}_{A'}=0.
\label{2.8}
\end{equation}
This equation yields the representation for the spin-tensor
$\acute{x}^{AA'}$ in the form\footnote{Representations (\ref{2.7}) and
(\ref{2.9}) imply $\dot{x}^2 = (\rho^\sigma /\rho^\tau )^2
\acute{x}^2$ and $\dot{x}\acute{x}= -(\rho^\sigma
/\rho^\tau)\acute{x}^2$. Hence, the determinant of the induced metric
on the null string worldsheet vanishes identically
($\dot{x}^2\acute{x}^2 - (\dot{x}\acute{x})^2 = 0$). This verifies
that the action principle (\ref{2.1}) provides a description of a null
string worldsheet.}
\begin{equation}
\acute{x}^{AA'}= o^A\bar{r}^{A'}+r^A\bar{o}^{A'},
\label{2.9}
\end{equation}
where condition $o_A r^A \not= 0$ is imposed on the spinor field $r^A$.

Action (\ref{2.1}) is invariant with respect to the following
transformations:
\begin{equation}
\tilde{o}^A = o^A, \,\,\,\,
\tilde{\iota}^A = {\iota}^A + uo^{A},
\label{2.10}
\end{equation}
and
\begin{equation}
\tilde{o}^A = \textrm{e}^v o^{A}, \,\,\,\,
\tilde{\iota}^A = \textrm{e}^{-v} \iota^{A}, \,\,\,\,
\tilde{\rho}^\mu = \textrm{e}^{-(v+\bar{v})}\rho^\mu
\label{2.11}
\end{equation}
with complex-valued functions $u(\xi)$ and $v(\xi)$.
Transformations (\ref{2.10}) and (\ref{2.11}) are usually referred to
as null rotations about the null direction corresponding to
$o^A$ and as boost-rotations, respectively. Expanding spinor
field $r^A$ in the basis ($o^A$,~$\iota^A$) as $r^A = p o^A + q\iota^A$,
where functions $p(\xi)$ and $q(\xi)$ are complex-valued,
representing $q$ in the polar form $q=|q|\exp(\i\varphi)$ and using
(\ref{2.9}) we find
\begin{equation}
\acute{x}^{AA'}=
 (p+\bar{p}) o^A\bar{o}^{A'} +
 |q|(\textrm{e}^{-\i\varphi} o^A\bar{\iota}^{A'} +
\textrm{e}^{\i\varphi}\iota^A\bar{o}^{A'}).
\label{2.12}
\end{equation}
Carrying out successive transformations (\ref{2.10}) and (\ref{2.11})
with parameters \\ $u= |q|^{-1}\bar{p}\exp(-\i\varphi )$ and
$v=-[\ln (\rho^\tau /e ) + \i\varphi ]/2$ we obtain
\begin{equation}
\dot{x}^{AA'} = o^A \bar{o}^{A'}
- \frac{\rho ^\sigma}{\rho ^\tau}\acute{x}^{AA'}, \,\,\,\,
\acute{x}^{AA'} = |q|(o^A \bar{\iota}^{A'} + \iota^A \bar{o}^{A'}).
\label{2.13}
\end{equation}
Here we have not put any tildes over the basis spinors, because motion
equations (\ref{2.5}) are invariant with respect to transformations
(\ref{2.10}) and (\ref{2.11}). Using the possibility of rescaling the
dyad element $\iota ^A$
\begin{equation}
\tilde{o}^A = o^{A}, \,\,\,\,
\tilde{\iota}^A = \lambda\iota ^{A}, \,\,\,\,
\tilde{\chi} = \lambda\chi,
\label{2.14}
\end{equation}
we incorporate factor $|q|$ into $\iota ^A$ and
write (\ref{2.13}) in the form
\begin{equation}
\dot{x}^{AA'} = o^A \bar{o}^{A'}
- \frac{\rho ^\sigma}{\rho ^\tau}\acute{x}^{AA'}, \,\,\,\,
\acute{x}^{AA'} = o^A \bar{\iota}^{A'} + \iota^A \bar{o}^{A'}.
\label{2.15}
\end{equation}
Using (\ref{2.15}) we represent the last equation in (\ref{2.5}) as
\begin{equation}
\partial_\mu (\rho^\mu o_A\bar{o}_{A'}) =
\i\kappa (\chi Q_{AB'}\bar{o}^{B'}\bar{o}_{A'} -
\bar{\chi}o_A Q_{BA'}o^B),
\label{2.16}
\end{equation}
where spin-tensor $Q^{AA'}$ corresponds to vector
$Q^a = \varepsilon^{abcd}\partial_{[b}B_{cd]}$ with
$3!\partial_{[a}B_{bc]}=
\varepsilon_{abcd}\varepsilon^{defg}\partial_{e} B_{fg}$
and we take $\varepsilon_{0123} = - \varepsilon^{0123} = 1$. For this
duality relation we shall call $Q^{AA'}$ the field strength of the
gauge potential $B_{ab}$.

Invariance of action (\ref{2.1}) under arbitrary reparametrizations of
the worldsheet implies (via the second Noether's theorem
\cite{Noether,Barbashov-Nesterenko1}) that the solutions of the motion
equations depend upon two arbitrary real-valued functions. We fix one
of them using the condition
\begin{equation}
\rho^\sigma =0.
\label{2.17}
\end{equation}
The null string equations of motion take the form
\begin{equation}
\dot{x}^{AA'} = o^A \bar{o}^{A'}, \,\,
\acute{x}^{AA'}= o^A \bar{\iota}^{A'} + \iota^A \bar{o}^{A'},
\,\,
(\rho^\tau o^A\bar{o}^{A'})\dot{} = {\cal F}^{AA'},
\label{2.18}
\end{equation}
where we have introduced the notation
\begin{equation}
{\cal F}_{AA'}= \i \kappa (\chi Q_{AB'}\bar{o}^{B'}\bar{o}_{A'} -
\bar{\chi}o_A o^B Q_{BA'}).
\label{2.19}
\end{equation}
By analogy with the case of a charged particle in a background
electromagnetic field we call ${\cal F}_{AA'}$ the Lorentz force
produced by the antisymmetric gauge potential $B_{ab}$. For the
purposes of future analysis we write the second equation of
system~(\ref{2.18}) as
\begin{equation}
\ddot{x}^{AA'} + ({\mathrm ln}|\rho^\tau|)\dot{}\,\dot{x}^{AA'} =
(\rho^\tau)^{-1}{\cal F}^{AA'}.
\label{2.20}
\end{equation}
Representation (\ref{2.18}) for $\dot{x}^{AA'}$ and $\acute{x}^{AA'}$
results in the Virasoro constraints
\begin{equation}
\dot{x}^2=0 \quad\mbox{and}\quad\dot{x}\acute{x} = 0
\label{2.21}
\end{equation}
appearing in the standard theory of null strings.
%------------------------------------------------------------------------------
%\newpage
\section{Null string identifies antisymmetric field as notoph}
Let us analyse null string equations of motion (\ref{2.18})~--~(\ref{2.20})
and constraints (\ref{2.21}). The presence of the Lorentz force
${\cal F}^{AA'}$ makes them different from that of the free null string
\cite{Karlhede-Lindstrom,Lizzi,Zheltukhin(2-3)}.

We start by showing that projections of ${\cal F}^{AA'}$ on
vectors $\dot{x}^{AA'}$ and $\acute{x}^{AA'}$ tangent to the null
string worldsheet vanish. Taking into account
normalization~(\ref{2.2}) we find the projections of the force
${\cal F}^{AA'}$ onto the spin-tensor basis elements $o_A\bar{o}{}_{A'}$,
$o_A\bar{\iota}{}_{A'}$, $\bar{o}_A\iota_{A'}$ and
$\iota_A\bar{\iota}_{A'}$
\begin{eqnarray}
%\fl 
{\cal F}^{AA'}o_A\bar{o}_{A'} = \dot{x}_{AA'}{\cal F}^{AA'} = 0, 
\,\,\,\,\,\,& &
{\cal F}^{AA'}\iota_A\bar{\iota}_{A'} \,=
-\i\kappa |\chi|^2 (o_A\bar{\iota}_{A'} -
\bar{o}_{A'}\iota_{A'})Q^{AA'},
\nonumber \\
%\fl 
{\cal F}^{AA'}o_A\bar{\iota}_{A'} =
\i\kappa |\chi|^2 o_A\bar{o}_{A'}Q^{AA'},& &
{\cal F}^{AA'}\iota_A\bar{o}_{A'} =
-\i\kappa |\chi|^2o_A\bar{o}_{A'}Q^{AA'}.
\label{3.1}
\end{eqnarray}

Therefore, we obtain
\begin{equation}
(o_A\bar{\iota}_{A'}+\iota_A\bar{o}_{A'}){\cal F}^{AA'} =
\acute{x}_{AA'}{\cal F}^{AA'} = 0.
\label{3.2}
\end{equation}
This shows that the force ${\cal F}^{AA'}$ possesses only two non-zero
projections onto the moving tetrad associated with each point of the
null string worldsheet. For the sake of brevity, we define the second
spacelike member of the moving tetrad by the condition
\begin{equation}
k^{AA'} = \i (o^A\bar{\iota}{}^{A'} - \bar{o}{}^{A'}\iota^A).
\label{3.3}
\end{equation}
Evidently $k^{AA'}$ is orthogonal to $\dot{x}^{AA'}$ and
$\acute{x}^{AA'}$. Then, subtracting the third and fourth equations in
(\ref{3.1}), the second non-zero projection of ${\cal F}^{AA'}$ is
given by
\begin{equation}
k_{AA'}{\cal F}^{AA'} = -2\kappa|\chi|^2o_A\bar{o}_{A'}Q^{AA'}.
\label{3.4}
\end{equation}

Now we are in a position to find the number of physical degrees of
freedom taking part in the interaction between the gauge field
$B_{ab}$ and the tensionless string. As we show below, this number is
equal to one.

First, we consider the case when the field strength $Q^{AA'}$ is
restricted by the condition
\begin{equation}
Q^{AA'}o_A\bar{o}_{A'} = Q^{AA'}\dot{x}_{AA'} = 0.
\label{3.5}
\end{equation}
In this case, the right-hand side of equation (\ref{3.4}) vanishes and
we are left with only one non-zero component of the Lorentz force,
${\cal F}^{AA'}\iota_A\bar{\iota}_{A'} =
-\kappa|\chi|^2k_{AA'}Q^{AA'}$. Taking into account the first pair
of equations in system~(\ref{2.18}) and equation~(\ref{3.3}) we can
write the general solution of equation (\ref{3.5}) for $Q^{AA'}$ as 
follows:
\begin{equation}
Q^{AA'}=s\dot{x}{}^{AA'} + r_{1}\acute{x}{}^{AA'} + r_{2}k^{AA'}.
\label{3.6}
\end{equation}
Here $s(\tau,\sigma)$, $r_1(\tau,\sigma)$ and $r_2(\tau,\sigma)$ are
arbitrary real-valued functions. Equation~(\ref{3.6}) shows that under
condition~(\ref{3.5}) the field strength $Q^{AA'}$ expands
into the triple of vectors $\dot{x}^{AA'}$, $\acute{x}^{AA'}$ and
$k^{AA'}$ of the moving tetrad. At a given point on the null string
worldsheet vectors $\dot{x}^{AA'}$ and $\acute{x}^{AA'}$ define
the flag plane associated with the flagpole $\dot{x}^{AA'}$ \cite{SST}. We
note in passing that this flag plane lies in the tangent plane of the
null string worldsheet. The vector $k^{AA'}$ can be obtained from
$\acute{x}^{AA'}$ by rotating the latter through a right angle about
the flagpole $\dot{x}^{AA'}$. This can also be achieved as a result
of performing transformations (\ref{2.11}) with parameter $v =
\i\pi/4$. Under such transformations $o^A\mapsto\exp{(\i\pi/4)}o^A$
and $\iota^A\mapsto\exp{(-\i\pi/4)}\iota^A$ which means that
$\acute{x}^{AA'}\mapsto k^{AA'}$. Rotations of the flag plane
through arbitrary angles sweep out a null hypersurface spanned by the
three mutually orthogonal vectors $\dot{x}^{AA'}$, $\acute{x}^{AA'}$
and $k^{AA'}$. The null character of this hypersurface follows
immediately from the null property of vector $\dot{x}^{AA'}$ which
is also normal to the hypersurface. Expansion (\ref{3.6}) means that
the field strength $Q^{AA'}$ restricted by condition (\ref{3.5})
belongs to this hypersurface at each point of the null string
worldsheet. Using the expression for ${\cal F}^{AA'}\iota_A\bar{\iota}_{A'}$
from system~(\ref{3.1}) together with representation (\ref{3.6}) and
equation~(\ref{2.20}) we obtain
\begin{equation}
\ddot{x}^{AA'}\iota_A\bar{\iota}{}_{A'} = -(\rho^\tau)^{-1}
|\chi|^2 (\dot{\rho}{}^\tau - 2\kappa r_2).
\label{3.7}
\end{equation}
We eliminate the right-hand side of this equation by fixing the second
gauge condition corresponding to the remaining reparametrization
symmetry \cite{Zheltukhin4} of the Virasoro constraints (\ref{2.21})
\begin{equation}
\tilde{\tau} = \tilde{\tau}(\tau,\sigma) \,\,\mbox{and}\,\,
\tilde{\sigma} = \tilde{\sigma}(\sigma).
\label{3.8}
\end{equation}
The required gauge choice is given by the following condition on the
worldsheet density component $\rho^\tau$:
\begin{equation}
\dot{\rho}{}^\tau - 2\kappa r_2 = 0
\,\, \Rightarrow \,\, \rho^\tau_{\rm fix}=\rho^\tau(0,\sigma) +
2\kappa\int_{0}^{\tau}r_{2}\,d\tau.
\label{3.9}
\end{equation}
The null string equations of motion in the background of antisymmetric
fields restricted by condition (\ref{3.5}) take the form
of free null string equations of motion
\begin{equation}
\ddot{x}^{a}=0.
\label{3.10}
\end{equation}
Therefore, external antisymmetric fields of the form
(\ref{3.6}) can be excluded from the interaction with the null string
by fixing suitable gauge conditions. The possibility of fixing such
gauge conditions corresponds to the local symmetries of the null string
action functional (\ref{2.1}). These local symmetries  describe the
reparametrization freedom of the null string worldsheet and
arbitrariness (\ref{2.10}) and (\ref{2.11}) inherent in the choice of
the dyad $o^A,\,\iota^A$. We conclude that antisymmetric background
fields $B_{ab}$ obeying condition~(\ref{3.5}) can be treated as pure gauge
fields in the present formalism.

In the general case the vector $Q^{AA'}$ is given by expansion in the
basis of moving tetrad
\begin{equation}
Q^{AA'} = s\dot{x}^{AA'} + r_1\acute{x}^{AA'} + r_{2}k^{AA'}
+ \phi\iota^A\bar{\iota}{}^{A'}.
\label{3.11}
\end{equation}
The tetrad is produced by addition of the null vector
$\iota^A\bar{\iota}{}^{A'}$ to the vectors $\dot{x}^{AA'}$,
$\acute{x}^{AA'}$, $k^{AA'}$. In the view of the statement proved
above the components $s$, $r_1$ and $r_2$ of the field strength
$Q^{AA'}$ does not contribute to the interaction with the null
string. Hence, only one component, $\phi$, of the field strength
$Q^{AA'}$ is essential for the interaction and cannot be excluded by
any gauge transformation. This implies that for the purposes of
interaction with a null string the behaviour of antisymmetric gauge
field $B_{ab}$ is classically equivalent to that of the real-valued
scalar field. The scalar field $\phi$ describes single gauge-invariant
physical degree of freedom which is naturally associated with the
notoph -- a massless particle of zero helicity
\cite{Ogievetskii-Polubarinov}. Our conclusion agrees with results of
Kalb and Ramond~\cite{Kalb-Ramond}. They found that tensor interactions
between closed tensile strings are effectively equivalent to those of
between a string and an antisymmetric background field $B_{ab}$
which has only one physical degree of freedom on its mass-shell. This
conclusion was derived in the framework of Wheeler-Feynman
action-at-a-distance approach \cite{Wheeler-Feynman} generalized to
the case of strings. Under such an approach physical fields appear as
secondary effective variables composed of worldline or worldsheet
coordinates of the interacting particles and strings\footnote{It is
interesting to observe at this point that compatibility of the
action-at-a-distance approach with global supersymmetry in Minkowski
spacetime has been recently shown in~\cite{Tugai-Zheltukhin} (see also
\cite{Alstine}).}. In spite of this fact Kalb and Ramond  had
proven that their results are equivalent to the results of the field
theory for the field $B_{ab}$.

In this connection let us briefly recall that the theory of an
antisymmetric gauge field $B_{ab}$ is characterized by the following
gauge symmetry:
\begin{equation}
{B'}_{ab}=B_{ab}+\partial_{a}\Lambda_{b}-\partial_{b}\Lambda_{a}.
\label{3.12}
\end{equation}
This is also one of the symmetries of the interacting null string
action (\ref{2.1}). The equation of motion for $B_{ab}$ is
\begin{equation}
\partial^{a}{\tilde F}_{a}=0,
\label{3.13}
\end{equation}
where ${\tilde F}_{a} =
\varepsilon^{abcd}\partial_{[b}B_{cd]}$. Defining a real-valued scalar
field $\tilde{\phi}$ by the relation
$\tilde{F}_a = \partial_a\tilde{\phi}$ we can represent~(\ref{3.13})
in the form
\begin{equation}
\Box
%\sqcup\hspace{-3.75mm}\sqcap  %This is substitution for \Box
\tilde{\phi} = 0.
\label{3.14}
\end{equation}
Thus, we see that antisymmetric field $B_{ab}$ has only one gauge-invariant 
degree of freedom which can be described by a real-valued massless scalar 
field.

Comparison of this description with the picture of the null string
interaction considered above shows that we can identify the vector
$Q^a$ and the corresponding real-valued scalar $\phi$ with
$\tilde{F^a}$ and $\tilde{\phi}$, respectively. This identification is
possible because of the fact that null string equations of motion
(\ref{2.18}) together with the remaining reparametrization symmetry
(\ref{3.8}) play a similar role to that of equation (\ref{3.13}) in the
theory of propagating antisymmetric field $B_{ab}$. In our analysis we
do not derive the Klein-Gordon equation for the field $\phi$. It was,
however, shown in~\cite{Callan} that wave equations for
Yang-Mills and other types of background fields which belong to the
string spectrum appear as conditions ensuring the absence of a conformal
anomaly on the quantum level. Since the quantum field theory of tensionless
(super)strings and (super)membranes is free of conformal
anomalies~\cite{Bandos-Zheltukhin}, the Klein-Gordon equation for the
field $\phi$ must appear in the quantum picture as a consequence of
this result.
%-----------------------------------------------------------------------------
%\newpage
\section{Analysis of equations of motion}
Having established the number of the physical degrees of freedom for
the  antisymmetric field interacting with the null string we can
now continue the analysis of the null string motion
equations~(\ref{2.18}).

Firstly, the representations of $\dot{x}^{AA'}$ and $\acute{x}^{AA'}$
must obey some compatibility conditions. Since $(\dot{x}^{AA'})\acute{}$
is equal to $(\acute{x}^{AA'})\dot{}$ the spinors $o^A$ and $\iota^A$
satisfy the  following relation:
\begin{equation}
\acute{o}^A\bar{o}^{A'} + o^A\acute{\bar{o}}{}^{A'} =
o^A\dot{\bar{\iota}}{}^{A'} + \dot{o}^A\bar{\iota}{}^{A'} +
\iota^A\dot{\bar{o}}{}^{A'} + \dot{\iota}^A\bar{o}{}^{A'}.
\label{4.1}
\end{equation}
Multiplying both sides of equation~(\ref{4.1}) by
$o_A\bar{o}_{A'}$ we get
\begin{equation}
\bar{\chi}\dot{o}^A o_A + \chi\dot{\bar{o}}{}^{A'}\bar{o}_{A'} = 0,
\label{4.2}
\end{equation}
which yields
\begin{equation}
\dot{o}^A o_A + \i\chi\psi = 0,
\label{4.3}
\end{equation}
where $\psi(\xi)$ is an arbitrary real-valued function. Allowing for
(\ref{2.2}), one can write
\begin{equation}
\dot{o}^A = \i\dot{\omega}o^A - \i\psi\iota^A.
\label{4.4}
\end{equation}
Here $\omega(\xi)$ is an arbitrary complex-valued function and the
multiplier $\i$ is introduced for future convenience. Substituting
(\ref{4.4}) into (\ref{4.1}) and projecting the resulted equation on
$o_A\bar{\iota}_{A'}$ we find
\begin{equation}
(\dot{\iota}^A - \i\dot{\bar{\omega}}\iota^A -
\acute{o}^A )o_A = 0
\label{4.5}
\end{equation}
and it follows that
\begin{equation}
\dot{\iota}^A - \i\dot{\bar{\omega}}\iota^A - \acute{o}^A  =
\hat{\mu} o^A.
\label{4.6}
\end{equation}
Substituting again (\ref{4.4}) and (\ref{4.6}) into (\ref{4.1}) we
obtain
\begin{equation}
\hat{\mu} = \i\tilde{\mu},
\label{4.7}
\end{equation}
where $\tilde{\mu}(\xi)$ is an arbitrary real-valued function. Thus,
compatibility condition (\ref{4.1}) results in the pair of equations for
the basis spinors $o^A$ and $\iota^A$
\begin{eqnarray}
\dot{o}^A & = & \i\dot{\omega}o^A - \i\psi\iota^A , \nonumber \\
\dot{\iota}^A & = & \i\tilde{\mu}o^A + \i\dot{\bar{\omega}}\iota^A
+ \acute{o}^{A}.
\label{4.8}
\end{eqnarray}

Secondly, we find that projection of both sides of the last
equation in~(\ref{2.18}), or equivalently, of equation~(\ref{2.20}) on
the flagpole direction $o_A\bar{o}_{A'}$ vanishes and its projection
onto $\acute{x}_{AA'}$ gives equation~(\ref{4.2}) by virtue of
using~(\ref{3.2}). Projection of (\ref{2.20}) onto the remaining
members of the tetrad, $k^{AA'}$ and $\iota_A\bar{\iota}_{A'}$ gives
\begin{eqnarray}
\bar{\chi}\dot{o}_{A}\iota^A + \chi\dot{\bar{o}}_{A'}\bar{\iota}^{A'}
& = & 0,
\nonumber \\
\bar{\chi}\dot{o}_A o^A - \chi\dot{\bar{o}}_{A'}\bar{o}^{A'} & = &
-2\i\kappa|\chi|^4(\rho^\tau_{\rm fix})^{-1}\phi,
\label{4.9}
\end{eqnarray}
where we took into account equations (\ref{3.1}) -- (\ref{3.4}) and
gauge condition~(\ref{3.9}). Using representation (\ref{4.8}) for
$\dot{o}_A$ in system (\ref{4.9}) we derive that $\omega(\xi)$ is a
real-valued function and $\psi(\xi)$ is given by
\begin{equation}
\psi = -\kappa|\chi|^2(\rho^\tau_{\mathrm fix})^{-1}\phi.
\label{4.10}
\end{equation}

We observe that in this approach the null string motion
equations~(\ref{2.20}) result in simple algebraic conditions which
reduce arbitrariness in the definition of the component $\omega$ in the
dyad decomposition of the spinor fields $\dot{o}^A$ and
$\dot{\iota}^A$ and identify the component $\psi$ with the notoph
field $\phi$. These components define the dynamics of the dyad moving
frame on the null string worldsheet. The same situation arises in
the Lund-Regge geometric approach to the dynamics of strings as
embedded surfaces in spacetime
\cite{Lund-Regge,Omnes,Zheltukhin6,Zheltukhin(5-7),Zheltukhin8,Barbashov-Nesterenko2}.
In this approach the string equations of motion are transformed into
algebraic relations for the first and second fundamental forms of the
string worldsheet, and the integrability conditions for the moving
frames associated with the worldsheet start playing a main dynamical role.
In
our case the integrability conditions are given by
equations~(\ref{4.8}). Allowing for reality of $\omega$ and the
interpretation of $\psi$ given above we can re-express them as
\begin{eqnarray}
\dot{o}^A & = & \i\dot{\omega}o^A - \i\psi\iota^A, \nonumber \\
\dot{\iota}^A & = & \i\tilde{\mu}o^A + \i\dot{\omega}\iota^A + \acute{o}^A.
\label{4.11}
\end{eqnarray}
From geometric point of view these equations are similar to the
well known Maurer-Cartan equations for the special case of isotropic
2-surfaces embedded into 4D~Minkowski spacetime in the background of
antisymmetric fields. The formulation of the Lund-Regge approach in
terms of Cartan's theory of moving frames  was considered in
\cite{Zheltukhin6,Zheltukhin(5-7),Zheltukhin8}, where a connection
between strings and 2D~Yang-Mills-Higgs theories was also established.
According to this gauge formulation the notoph field
finds interpretation as some additional component of the worldsheet
Yang-Mills-Higgs field.

We can further simplify equations~(\ref{4.11}) by choosing spinor
variables $u^{A}(\tau,\sigma)$ and $v^{A}(\tau,\sigma)$
\begin{equation}
o^A = e^{\i\omega}u^A, \,\,\,\, \iota^A = e^{\i\omega}v^A,
\label{4.12}
\end{equation}
where the scalar products of new spinors are
\begin{equation}
u_Av^A = e^{-2\i\omega}\chi, \,\,\,\,
\bar{u}_{A'}\bar{v}^{A'} = e^{2\i\omega}\bar{\chi}.
\label{4.13}
\end{equation}
Using these spinors we rewrite equations~(\ref{4.11}) in the form
\begin{eqnarray}
\dot{u}^A & = & -\i\psi v^A, \nonumber \\
\dot{v}^A & = & \i\mu u^A + \acute{u}^A,
\label{4.14}
\end{eqnarray}
where $\mu = \tilde{\mu} + \acute{\omega}$ is a real-valued function of
$\tau$ and $\sigma$. If we re-express $v^A$ through  $\dot{u}^A$ using the
first equation in system~(\ref{4.14}) and substitute the result into
the second equation then we obtain a homogeneous second-order partial
differential equation for the basis spinor $u^A$
\begin{equation}
\ddot{u}_{A} - (\mbox{ln}|\psi|)\dot{}\,\dot{u}_{A} +
\i\psi\acute{u}_A - \mu\psi u_A = 0.
\label{4.15}
\end{equation}
Solutions of this equation define the second basis spinor $v^A$ via the
first equation in system (4.14). Equation~(\ref{4.15}) must be accompanied
by corresponding initial data and appropriate  periodicity
conditions in the case of closed strings. Introducing real $\zeta_A$
and imaginary $\eta_A$ parts of the spinor field $u_A$
\begin{equation}
u_A = \zeta_A + \i\eta_A
\label{4.16}
\end{equation}
together with linear differential operators $L_\tau$ and $L_\sigma$
\begin{equation}
L_\tau = \partial_\tau^2 -
(\mbox{ln}|\psi|)\dot{}\,\partial_\tau - \mu, \,\,\,\,
L_\sigma = \psi\partial_\sigma
\label{4.17}
\end{equation}
allows to write the null string equations of motion in the matrix form,
\begin{equation}
\left(
\begin{array}{rr}
L_\tau   & - L_\sigma \\
L_\sigma & L_\tau
\end{array}
\right)
\left(
\begin{array}{l}
\zeta_A \\
\eta_A
\end{array}
\right) = 0.
\label{4.18}
\end{equation}
Equation~(\ref{4.15}), or its matrix representation (\ref{4.18}), is
final equation which completely determines evolution of the tensionless
string in the background of antisymmetric fields. Solutions of
these equations will be studied in another paper.
%------------------------------------------------------------------------------
%\newpage
\section{Conclusion}
We studied the dynamics of the null string embedded in an arbitrary
background of antisymmetric fields and formulated a condition for the
absence of interaction with such fields. According to this
condition the null string identifies the field strength of the
antisymmetric field as the notoph -- a massless particle of zero
helicity. We show that non-linear equations of motion of the null string
in background antisymmetric fields can be reduced to algebraic
constraints and integrability conditions for 2-component spinor
representation of $\dot{x}^{AA'}$ and $\acute{x}^{AA'}$, and the
integrability conditions play the role of dynamical equations. The total
effect of the null string interaction with such fields is contained
in a single derivation coefficient of the decomposition of
$\dot{o}^A$ into the dyad basis. This derivation coefficient is
given by projection $\dot{o}^A$ on the dyad basis spinor $o_A$, and we
interpret this coefficient as the notoph field.  In the case of
free null string this derivation coefficient
vanishes. Therefore, we can detect a notoph field with the aid of a
tensionless string. Under the gauge field theory
interpretation of the Lund-Regge geometric approach suggested in
\cite{Zheltukhin6,Zheltukhin(5-7)} these derivation coefficients play
the role of the components of the 2D~Yang-Mills-Higgs field associated with
the holonomy group of the string worldsheet. Thus from the
point of view of 2D~worldsheet physics the appearance of the notoph
field means the presence of an additional component of the 2D~Yang-Mills
field associated with the null string embedded into the background of
antisymmetric fields.

We also find that equations of motion defining the dynamics of the null
string can be transformed into a linear homogeneous system of second-order 
partial differential equations. Hence, 2-component spinor formalism 
provides a natural geometric framework for reduction of non-linear null 
string equations of motion in the background of antisymmetric fields to 
linear ones. At this point, it would be interesting to consider 
generalizations of these results to the case of tensionless p(D)-branes 
propagating in external antisymmetric fields of higher rank in higher 
dimensions of spacetime.  
%------------------------------------------------------------------------------
%\newpage
\section{Acknowledgements}
This work is supported, in part, by Dutch Government and INTAS Grants
No~94-2317, No~93-633-ext and No~93-127-ext. The first author~(KI)
would like to acknowledge the Dervorguilla Postgraduate Scholarship in the 
Sciences awarded by Balliol College. He is also grateful to Profs~R~Penrose 
and Yu~P~Stepanovsky for fruitful discussions. The second author~(AAZ) 
wishes to acknowledge financial support by the Government Fund for 
Fundamental Research of the Ministry of Research and Technology of Ukraine, 
a High Energy Physics Grant awarded by the ministry and a Grant of Royal 
Swedish Academy of Sciences. He is grateful to Prof~U~Lindstr\"{o}m.
%------------------------------------------------------------------------------
\newpage
%\section*{References}

%------------------------------------------------------------------------------

\begin{references}
\bibitem{Veneziano} Veneziano G 1996 {\it String Gravity and Physics
      at the Planck Energy Scale (Erice), NATO ASI Series} C {\bf 476}
      ed~N~S$\acute{\mbox{a}}$nchez and A~Zichichi p~285
\bibitem{deVega-Sanchez} de~Vega H J and S$\acute{\mbox{a}}$nchez N
      1996 {\it String Gravity and Physics at the Planck Energy Scale
      (Erice), NATO ASI Series} C {\bf 476}
      ed~N~S$\acute{\mbox{a}}$nchez and A~Zichichi p~11
\bibitem{Antoniadis} Antoniadis I and Obers N A 1994
      {\it Nucl.~Phys.} B {\bf 243} 639
\bibitem{Roshchupkin-Zheltukhin} Roshchupkin S N and Zheltukhin A A 1995
      {\it Class.~Quantum~Grav.} {\bf 12} 2519
\bibitem{Kar(1-2)} Kar S 1996 {\it Phys.~Rev.} D {\bf 53} 6842 \\
%  \nonum  
	 Kar S 1996 {\it Phys.~Rev.} D {\bf 54} 6408
\bibitem{Dabrowski-Larsen} Dabrowski M P and Larsen A L {\it Preprint}
      hep-th/9610243
\bibitem{Schild} Schild A 1977 {\it Phys.~Rev.} D {\bf 16} 1722
\bibitem{Zheltukhin1} Zheltukhin A A 1996 {\it Class.~Quantum~Grav.}
      {\bf 13} 2357
\bibitem{deVega-Giannakis-Nicolaidis} de~Vega H J,
      Giannakis I and Nicolaidis A 1995 {\it Mod.~Phys.~Lett.} A
      {\bf 10} 2479
\bibitem{Lousto-Sanchez} Lousto C O and S$\acute{\mbox{a}}$nchez N 1996
      {\it Phys.~Rev.} D {\bf 54} 6399
\bibitem{Karlhede-Lindstrom} Karlhede A and Lindstr$\ddot{\mbox{o}}$m U
      1986 {\it Class.~Quantum~Grav.} {\bf 3} L73
\bibitem{Lizzi}  Lizzi F, Ray B, Sparano G and Srivastava A
      1986 {\it Phys.~Lett.} B {\bf 182} 112
\bibitem{Zheltukhin(2-3)} Zheltukhin A A 1987 {\it Sov.~Phys.-JETP~Lett.}
      {\bf 46} 262 \\
%  \nonum  
	 Zheltukhin A A 1988 {\it Sov.~J.~Nucl.~Phys.} {\bf 48} 375
\bibitem{Zheltukhin4} Zheltukhin A A 1989 {\it Phys.~Lett.} B {\bf 233}
      112 \\
%  \nonum 
	 Zheltukhin A A 1990 {\it Sov.~J.~Nucl.~Phys.} {\bf 51} 1504
\bibitem{Town1} Townsend P K 1994 {\it Phys.~Lett.} B {\bf 277} 275
\bibitem{Has-Lin} Hassani~S, Lindstr\"{o}m~U and Von~Unge~R  1994 {\it
      Class.~Quant.~Grav.} {\bf 11} L79
\bibitem{De-Ga-Ho-So} Delduc~F, Galperin~A, Howe~P and Sokatchev~E
      1992 {\it Phys.~Rev.} D {\bf 46} 714
\bibitem {Bandos-Zheltukhin0} Bandos I A and Zheltukhin A A  1993
      {\it Fortschr.~Phys.} B {\bf 41} 619
\bibitem{Hughston-Shaw} Hughston L P and Shaw W T 1990 {\it LMS Lecture
      Notes} {\bf 156} p~218
\bibitem{Bandos-Zheltukhin} Bandos I A and Zheltukhin A A 1991
      {\it Theor.~Math.~Phys.}  {\bf 88} 358 \\
%  \nonum 
	 Bandos I A and Zheltukhin A A 1991 {\it Phys. Lett.} B
      {\bf 261} 245
\bibitem{SST} Penrose~R and Rindler~W 1984 {\it Spinors and
      Space-Time} vols~1~\&~2 (New York: CUP)
\bibitem{(1-2)Vol-Zhe} Volkov D V and Zheltukhin A A 1988
      {\it Sov.~Phys.~-~JETP~Lett.} {\bf 48} 61 \\
%  \nonum 
	 Volkov D V and Zheltukhin A A 1989 {\it Lett.~Math.~Phys.}
      {\bf 17} 141
\bibitem{3Vol-Zhe} Volkov D V and Zheltukhin A A 1990
      {\it Nucl.~Phys.} B {\bf 335} 6
\bibitem{Sor-Tk-Vol} Sorokin~D, Tkach~V and Volkov~D~V 1989
      {\it Mod.~Phys.~Lett.} A {\bf 4} 901
\bibitem{Sor-Tk-Vol-Zhe} Sorokin~D, Tkach~V, Volkov~D~V and
      Zheltukhin~A~A 1989 {\it Phys.~Lett.} B {\bf 216} 901
\bibitem{Ho-Town} Howe P S and Townsend P K 1989 {\it Phys.~Lett.} B
      {\bf 259} 285
\bibitem{Gal-Soc} Galperin A and Sokatchev E 1992 {\it Phys.~Rev.} D
      {\bf 46} 714
\bibitem{Iva-Kap} Ivanov E and Kapustnikov A 1991 {\it Phys.~Lett.} B
      {\bf 267} 175
\bibitem{Chi-Pa} Chikalov V and Pashnev A 1993 {\it Mod.~Phys.~Lett.} A
      {\bf 8} 901
\bibitem{Ber-Sez} Bergshoeff E and  Sezgin E 1994 {\it Nucl.~Phys.} B
      {\bf 266} 312
\bibitem{Berkov} Berkovitz N 1989 {\it Phys.~Lett.} B {\bf 232} 184 \\
%  \nonum 
	 Berkovitz N 1990 {\it Phys.~Lett.} B {\bf 241} 497
\bibitem{Jensen-Lindstrom} Jensen B and Lindstr\"{o}m U
      1997 {\it Phys.~Lett.} B {\bf 398} 83
\bibitem{Ogievetskii-Polubarinov} Ogievetskii V I and Polubarinov I V
      1967 {\it Sov.~J.~Nucl.~Phys.} {\bf 4} 156
\bibitem{Kalb-Ramond} Kalb M and Ramond P 1974 {\it Phys.~Rev.} D
      {\bf 9} 2273
\bibitem{Lund-Regge} Lund F and Regge T 1976 {\it Phys.~Rev.} D14 1524
\bibitem{Omnes} Omnes R 1979 {\it Nucl.~Phys.} B {\bf 149} 269
\bibitem{Zheltukhin6} Zheltukhin A A 1981 {\it Sov.~J.~Nucl.~Phys.}
      {\bf 33} 1723 \\
%  \nonum 
	 Zheltukhin A A 1981 {\it Lett. Math. Phys.}{\bf 5} 213
\bibitem{Zheltukhin(5-7)} Zheltukhin A A 1982 {\it Sov.~J.~Math.~Phys.}
      {\bf 52} 73 \\
%  \nonum 
	 Zheltukhin A A 1982 {\it Phys. Lett.} B {\bf 116} 147
\bibitem{Zheltukhin8} Zheltukhin A A 1983 {\it Sov.~J.~Math.~Phys.}
      {\bf 56} 230
\bibitem{Barbashov-Nesterenko1} Barbashov B M, Nesterenko V V and
      Dumbrajs T Yu 1990 {\it Introduction to the Relativistic String
      Theory} (Singapore: World Scientific)
\bibitem{Noether} Noether E 1918 {\it Nach. v. d. Ges. d. Wiss. zu
      G$\ddot{\mathrm{o}}$ttingen} 235
\bibitem{Barbashov-Nesterenko2} Barbashov B M and Nesterenko V V 1983
      {\it Fortschr. Phys.} {\bf 31} 535
\bibitem{Wheeler-Feynman} Wheeler J A and Feynman R 1949 {\it
      Rev.~Mod.~Phys.} {\bf 21} 425
\bibitem{Alstine} Van~Alstine~P and Crater~H 1986 {\it Phys.~Rev.} D
      {\bf 33} 1037
\bibitem{Tugai-Zheltukhin} Tugai V and Zheltukhin A A 1995 {\it
      Phys.~Rev.} D {\bf 51} R3997 \\
%  \nonum 
	 Tugai V and Zheltukhin A A 1996 {\it Phys.~Rev.} D {\bf 54} 4160 \\
%  \nonum 
	 Tugai V and Zheltukhin A A {\it Preprint} hep-th/9706114
\bibitem{Callan} Callan~C~G, Friedan~D, Martinec~E~J and Perry~M~J
      1985 {\it Nucl. Phys.} B {\bf 262} 593
\end{references}
\end{document}